\documentclass[english,pra,showpacs,twocolumn,amsmath]{revtex4}
\usepackage{graphicx}
\usepackage{simplewick}
\usepackage{latexsym}

\begin{document}

\title{Model for inner structure and mass spectrum of charged leptons}

\author{Vladimir I. Kruglov}

\affiliation{Physics Department, The University of Auckland, Private Bag 92019, Auckland, New Zealand}

\affiliation{Corresponding author: v.kruglov@auckland.ac.nz}

\begin{abstract}
We present a model where the lepton masses are the eigenvalues of relativistic nonlinear field equations. The eigenfunctions correspond in the model to lepton states with inner structure. In this picture the self-interaction leads to the bound state of electron described by the ground state solution.
The variational approach based on minimization of the mass or the energy of the ground state solution is developed. The analytical solution yields the relation between the coupling constant $G_l$ of the self-interaction and the electron mass as $m_e\simeq 2.3\sqrt{\hbar c/G_l}$. This solution also leads to a finite `radius' of the electron which is about $1.9\lambda_e$ where $\lambda_e$ is the Compton length of electron. 
\end{abstract}


\maketitle

\section{Introduction}

The charged lepton pole masses obey the Koide relation \cite{Koi,Koid} with remarkable precision. All three charged lepton pole masses are given to $O(10^{-5})$ accuracy by a more general equation found independently in the context of a resurrected preon model. This equation for lepton masses has the form $(m_{e},m_{\mu},m_{\tau})=\tilde{m}(1+\sqrt{2}\cos {\theta_n)^2}$, where $\theta_n=2\pi n/3+2/9$ with the generation number $n=1,2,3$ and $\tilde{m}$ is the empirical mass constant \cite{Bran}. 
This empirical mass formula admits a geometrical interpretation in a model proposed by Dirac \cite{Dir} and extended in Ref.  \cite{Gol,Gold}. 

The modernized Dirac model also supports the proposal \cite{Ger} that leptons and quarks of the same generation are identical in size and shape. It is assumed in this approach that leptons and quarks are of finite size and have scale-independent pole masses that are the eigenvalues of some positive-definite self-adjoint operator, as originally proposed in Ref. \cite{Dir}. Moreover, analysis of the quark pole mass data and earlier empirical studies \cite{Ros} yields the extension of the above mass formula to quarks and Majorana neutrinos \cite{Ger}.

The extension of an empirical charged lepton mass relation to the neutrino sector is proposed in Ref. \cite{Rod}. It is assumed that the neutrinos are Majorana particles whose mass presumably originates from the seesaw mechanism \cite{Min,Moh}. This means that the light neutrino masses are related to the masses in the Dirac and heavy Majorana mass matrix. Another $A_4$ extension of the Standard Model (SM) has been proposed in Ref. \cite{King} leading to the quark-lepton mass relation. This model yields quark masses, mixing angles and CP violation by numerical fit. 

In the present Letter we propose a model describing the masses of leptons as eigenvalues of the nonlinear relativistic field equations. The model is a generalization of Quantum Electro-Dynamics (QED) by a real scalar $\Theta$-field. The relativistic extension of the QED Lagrangian by a scalar field leads to self-interaction of the fermion field. Moreover, this extended Lagrangian is invariant under local phase transformations described by the ${\rm U}(1)_Q$ symmetry group, i.e.\ it is gauge-invariant.

It is assumed in this model that the system of eigenfunctions of the fermion field correspond to a family of charged leptons with the masses given by appropriate  eigenvalues. These masses have the form:  $m_k=E_k/c^2$ where $E_k$ is the eigen-energy of the corresponding bound fermion state. 

In Section 2, we introduce   the Lagrangian of the model and derive the set of nonlinear equations describing the lepton family as the eigenfunctions with corresponding masses. In Sections 3, we  
describe the variational ground state solution based on trial functions which are connected with the spherical spinors. We present the main results in Section 4. In particular, we estimate in this section the lepton's coupling constant describing self-interaction. Finally, we summarize the main results of our work in Section 5. In the last section we also discuss the connection between the model presented here and the SM.

 \section{Model for lepton spectrum}
 
In our model the masses of leptons are the eigenvalues of the nonlinear relativistic field equations. The model is a generalization of QED by a real scalar $\Theta$-field. The relativistic extension of the QED Lagrangian by a real scalar field leads to self-interaction of the fermion field. Moreover, this Lagrangian is invariant under local phase transformations described by a ${\rm U}(1)_Q$ symmetry group.

We use in this section the units $\hbar=c=1$ and the metric signature $\eta_{\mu\nu}=(+,-,-,-)$. The gauge-invariant Lagrangian for charged leptons in our model is given by
\begin{multline}
{\cal L} =  \bar{\Psi} \gamma^{\mu}(i\partial_{\mu}-eA_{\mu})\Psi-(m+\Theta)\bar{\Psi}\Psi \\
-\frac{1}{16\pi}F_{\mu\nu}F^{\mu\nu}-\frac{1}{2\kappa}(\partial_{\mu}\Theta)(\partial^{\mu}\Theta)
-\frac{1}{\kappa}{\cal Q}(\Theta),~~
\label{1}
\end{multline}
where $F_{\mu\nu}=\partial_{\mu}A_\nu-\partial_{\nu}A_\mu$ and the Lorentz gauge $\partial_{\mu}A^{\mu}=0$ is assumed for the vector $A^{\mu}$ describing the electromagnetic field. The lepton charge is negative ($e<0$) in our notation. The Lagrangian in Eq.\ (\ref{1}) yields the system of equations for the lepton field $\Psi$, the four-vector $A^{\mu}$ and the real scalar field $\Theta$ as
\begin{equation}
\gamma^{\mu}(i\partial_{\mu}-eA_{\mu})\Psi-(m+\Theta)\Psi=0,
\label{2}
\end{equation}
\begin{equation}
\Box A^{\mu}-4\pi e\bar{\Psi}\gamma^{\mu}\Psi=0,
\label{3}
\end{equation}
\begin{equation}
\Box \Theta-{\cal Q}'(\Theta)-\kappa\bar{\Psi}\Psi=0,
\label{4}
\end{equation}
where $\Box=\partial_{\mu}\partial^{\mu}$, $\bar{\Psi}$ is the Dirac adjoint and ${\cal Q}'$ is the derivative [${\cal Q}'(\Theta)=d{\cal Q}/d\Theta$]. The coupling parameter $\kappa$ and normalization condition have the forms 
\begin{equation}
\kappa=4\pi G_lm^n,~~~\int\Psi^{\dag}\Psi d{\bf x}=1~.
\label{5}
\end{equation}
Here the parameter $n$ is some integer number which we choose later as $n=2$. The time-dependent lepton field is given by
\begin{equation}
 \Psi({\bf x},t)=\left( \begin{array}{c} \phi({\bf x},t) \\ \chi({\bf x},t) \end{array} \right )=\exp(-iEt)\left( \begin{array}{c} \phi({\bf x}) \\ \chi({\bf x}) \end{array} \right ),
 \label{6}
\end{equation}
with standard notations for spinors $\phi$ and $\chi$.

The four-vector has the form $A^{\mu}=(\Phi,{\bf A})$ and the magnetic field is given by ${\bf B}={\rm rot}{\bf A}$. Using the standard representation we can write Eq.\ (\ref{2}) as

\begin{equation}
(E-m)\phi=-\vec{\sigma}(i\nabla+e{\bf A})\chi+(\Theta+{\rm U})\phi,
\label{7}
\end{equation}
\begin{equation}
(E+m)\chi=-\vec{\sigma}(i\nabla+e{\bf A})\phi-(\Theta-{\rm U})\chi~,
\label{8}
\end{equation}
where ${\rm U}=e\Phi$ is the potential. Eqs.\ (\ref{3}) and (\ref{4}) yield the equations for the potential ${\rm U}$ and the scalar field $\Theta$ given by
\begin{equation}
\nabla^2 {\rm U}=-4\pi e^2(\phi^{\dag}\phi+\chi^{\dag}\chi),
\label{9}
\end{equation}
\begin{equation}
\nabla^2 \Theta+{\cal Q}'(\Theta)=-4\pi G_lm^n(\phi^{\dag}\phi-\chi^{\dag}\chi).
\label{10}
\end{equation}
The normalization condition for spinors follows from Eq.\ (\ref{5}) as $\int(\phi^{\dag}\phi+\chi^{\dag}\chi)d{\bf x}=1$.
We neglect later the small correction connected with the vector potential ${\bf A}$.

The spinors $\phi$ and $\chi$ can be written in the form
\begin{equation}
\phi({\bf x})=u(r)\Omega_{jlM}({\bf n}),~~~\chi({\bf x})=iv(r)\Omega_{jl'M}({\bf n}),
\label{11}
\end{equation}
where $l=j\pm 1/2$, $l'=j\mp 1/2$ and hence $l+l'=2j$, and $l-l'=\pm 1$. Here $\Omega_{jlM}$ and $\Omega_{jl'M}$ are the spherical spinors and ${\bf n}={\bf x}|{\bf x}|^{-1}$ is the unit vector. The spinors $\Omega_{jlM}({\bf n})$ can be normalized as
\begin{equation}
\int\Omega_{jlM}^{\dag}({\bf n})\Omega_{j'l'M'}({\bf n})d{\cal O}=\delta_{jj'}\delta_{ll'}\delta_{MM'},
\label{12}
\end{equation}
where $d{\cal O}=\sin\theta d\theta d\phi$. We also note that spherical spinors satisfy the relations
\begin{equation}
(\vec{\sigma}{\bf n})\Omega_{jlM}({\bf n})=-\Omega_{jl'M}({\bf n}),~(\vec{\sigma}{\bf n})\Omega_{jl'M}({\bf n})=-\Omega_{jlM}({\bf n}).
\label{13}
\end{equation}
The derivatives in Eq.\ (\ref{7}) and (\ref{8}) can be determined in spherical coordinates by the equations:
\begin{equation}
(\vec{\sigma}\nabla)\Omega_{jlM}({\bf n})=-(1+\beta)\Omega_{jl'M}({\bf n})\frac{1}{r},
\label{14}
\end{equation}
\begin{equation}
(\vec{\sigma}\nabla)\Omega_{jl'M}({\bf n})=-(1-\beta)\Omega_{jlM}({\bf n})\frac{1}{r},
\label{15}
\end{equation}
where the parameter $\beta$ is given by
\begin{equation}
\beta=l(l+1)-j(j+1)-\frac{1}{4}.
\label{16}
\end{equation}
We note that $\beta=-j-1/2$ for $l=j-1/2$, hence for the ground state we have $l=0$ and $j=1/2$. This yields for the ground state the parameter $\beta=-1$. 

We assume that the energy of the ground state is given by
$E=m$ (in standard units $E=mc^2$). Using the above equations for spherical spinors we can write Eqs.\ (\ref{7}) and (\ref{8}) for the ground state with $\beta=-1$ and $E=m$ as
\begin{equation}
\frac{du}{dr}=(2m+\Theta-{\rm U})v,
\label{17}
\end{equation}
\begin{equation}
\frac{dv}{dr}+\frac{2v}{r}=(\Theta+{\rm U})u.
\label{18}
\end{equation}
For the ground state
($l=0$ and $j=1/2$) Eqs.\ (\ref{9}) and (\ref{10}) become 
\begin{equation}
\frac{d^2}{dr^2}{\rm U}+\frac{2}{r}\frac{d}{dr}{\rm U}=-e^2(u^2+v^2),
\label{19}
\end{equation}
\begin{equation}
\frac{d^2}{dr^2}\Theta+\frac{2}{r}\frac{d}{dr}\Theta+{\cal Q}'(\Theta)=-G_lm^n(u^2-v^2).
\label{20}
\end{equation}
The normalization condition for the functions $u(r)$ and $v(r)$ following from Eq.\ (\ref{5}) is
\begin{equation}
\int_0^{+\infty}[u(r)^2+v(r)^2]r^2dr=1.
\label{21}
\end{equation}
In the following sections we choose the exponential parameter in Eqs.\ (\ref{5}) and (\ref{20}) to be $n=2$.
This leads to the interaction or coupling of two particles being proportional to their masses.
 
We assume that the eigenfunctions of the system nonlinear integro-differential Eqs.\ (\ref{17}-\ref{21}) correspond to a family of charged leptons with the masses given by the eigenvalues. These masses have the form $m_k=E_k/c^2$ (in standard units) with $m_1<m_2<m_3<...$ where $E_k$ ($k=1,2,3,...$) is the eigenenergy of the corresponding bound fermion state. 
The lepton masses are given by the three smallest eigenvalues: $m_1=m_e$, $m_2=m_{\mu}$, $m_3=m_{\tau}$ and the corresponding eigenfunctions evidently describe the
inner structure of free $e$, $\mu$, and $\tau$ leptons. Thus, the electron inner structure can be  described by the ground state of the system of Eqs.\ (\ref{17})--(\ref{21}), and $\mu$ and $\tau$ leptons correspond to excited states of this system of equations. This picture is consistent with the fact that the electron is stable but $\mu$ and $\tau$ particles are not stable.

\section{Ground state solution}

In this section, using the variational method for the ground state solution, we solve the system of equations given by Eqs.\ (\ref{17})--(\ref{21}). The variational method yields the minimal eigenvalue $m_1=m_e$ which is the mass of electron. The bound state or inner structure for a free electron is given by the ground state solution. We note that the electron (like other leptons) does not have any inner structure in the standard theories, including QED and SM.

We show below that the variational solution yields an inner structure for the electron with a characteristic size or  radius about two Compton lengths.
In this section we use standard units together with the definitions
\begin{equation}
\mu=\frac{mc}{\hbar},~~~\gamma=\frac{G_lm^n}{\hbar c},~~~\alpha=\frac{e^2}{\hbar c},
\label{22}
\end{equation}
where $\alpha\simeq 1/137$. Eqs.\ (\ref{17}) and (\ref{18}) can be written as
\begin{equation}
\frac{du}{dr}=(2\mu+\tilde{\Theta}-\tilde{{\rm U}})v,
\label{23}
\end{equation}
\begin{equation}
\frac{dv}{dr}+\frac{2v}{r}=(\tilde{\Theta}+\tilde{{\rm U}})u,
\label{24}
\end{equation}
where $\tilde{\Theta}=\Theta/(\hbar c)$ and $\tilde{{\rm U}}={\rm U}/(\hbar c)$.
We assume below that ${\cal Q}(\Theta)\equiv 0$. In this case the integration of Eqs.\ (\ref{19}) and (\ref{20}) with appropriate boundary conditions leads to the potential $\tilde{{\rm U}}(r)$ and the field $\tilde{\Theta}(r)$ being given by
\begin{equation}
\tilde{{\rm U}}(r)=\alpha\int_r^{+\infty}\frac{ds}{s^2}\int_0^s[u^2(r')+v^2(r')]r'^2dr',
\label{25}
\end{equation}
\begin{equation}
\tilde{\Theta}(r)=\gamma\int_r^{+\infty}\frac{ds}{s^2}\int_0^s[u^2(r')-v^2(r')]r'^2dr'.
\label{26}
\end{equation}
Thus the full system of equations consists of Eqs.\ (\ref{23})--(\ref{26}) and the normalization condition given by Eq.\ (\ref{21}). 

The variational ground state solution of this system of nonlinear integro-differential equations can be found by appropriate trial functions approximating the eigenfunctions $u(r)$ and $v(r)$ with the minimal eigenvalue given by $E=mc^2$ which yields the minimal parameter $\mu$. We choose trial functions for the Eqs.\ (\ref{23})--(\ref{26}) in the form
\begin{equation}
u(r)=A\exp\left(-\frac{1}{2}\sigma r\right),~~~v(r)=B\exp\left(-\frac{1}{2}\sigma r\right).
\label{27}
\end{equation}
In this case Eqs.\ (\ref{25}) and (\ref{26}) yield 
\begin{equation}
\tilde{{\rm U}}(r)=-\alpha(A^2+B^2)F(r),
\label{28}
\end{equation}
\begin{equation}
\tilde{\Theta}(r)=-\gamma(A^2-B^2)F(r),
\label{29}
\end{equation}
where the function $F(r)$ is given by
\begin{equation}
F(r)=\frac{1}{\sigma^2}e^{-\sigma r}-\frac{2}{\sigma^3r}(1-e^{-\sigma r}).
\label{30}
\end{equation}
The normalization condition Eq.\ (\ref{21}) leads to the coefficients $A$ and $B$ of the trial functions 
being related by
\begin{equation}
A^2+B^2=\frac{\sigma^3}{2}.
\label{31}
\end{equation}
Multiplying Eqs.\ (\ref{23}) and (\ref{24}) by $u$ and $v$ respectively and integrating over an infinite volume one can write 
\begin{equation}
\int\frac{du}{dr}udV=2\mu \int uvdV+R\int FuvdV,
\label{32}
\end{equation}
\begin{equation}
\int\frac{dv}{dr}vdV+2\int \frac{v^2}{r}dV=N\int FuvdV,
\label{33}
\end{equation}
where $dV=4\pi r^2dr$ and the constants $R$ and $N$ are
\begin{equation}
R=\gamma(A^2-B^2)+\frac{\alpha\sigma^3}{2},~~~N=\gamma(A^2-B^2)-\frac{\alpha\sigma^3}{2}.
\label{34}
\end{equation}
Eqs.\ (\ref{32}) and  (\ref{33}) with the trial functions of Eq.\ (\ref{27}) yield two equations which can be written as
\begin{equation}
B=-aA,~~~A=-bB,
\label{35}
\end{equation}
where the parameters $a$ and $b$ are
\begin{equation}
a=-\frac{5\gamma}{4\sigma^3}(A^2-B^2)-\frac{5\alpha}{8},
\label{36}
\end{equation}
\begin{equation}
b=\frac{4\mu}{\sigma}+\frac{5\gamma}{4\sigma^3}(A^2-B^2)-\frac{5\alpha}{8}.
\label{37}
\end{equation}
From the above equations follow the two relations
\begin{equation}
a+b=\frac{4\mu}{\sigma}-\frac{5\alpha}{4},~~~ab=1,
\label{38}
\end{equation}
which lead to a quadratic equation for the parameter $a$:
\begin{equation}
a^2-\nu a+1=0,~~~\nu=\frac{4\mu}{\sigma}-\frac{5\alpha}{4}.
\label{39}
\end{equation}
This equation has two different solutions
\begin{equation}
a=\frac{\nu}{2}\pm\frac{1}{2}\sqrt{\nu^2-4},
\label{40}
\end{equation}
with the condition $\nu^2\geq 4$. Eqs.\ (\ref{31}) and (\ref{35}) also lead to results for the coefficients $A$ and $B$ as
\begin{equation}
A^2=\frac{\sigma^3}{2(a^2+1)},~~~B^2=\frac{\sigma^3a^2}{2(a^2+1)}.
\label{41}
\end{equation}
Combining Eq.\ (\ref{36}) and Eq.\ (\ref{41}) we find another quadratic equation for the parameter $a$ which can be written as
\begin{equation}
(5\gamma-8\nu)a^2-5\alpha\nu a-5\gamma=0.
\label{42}
\end{equation}
One can neglect in this equation the small constant $\alpha\simeq 1/137$; the equation is then 
\begin{equation}
5\gamma(a^2-1)=8\nu a^2,~~~\nu=\frac{4\mu}{\sigma}.
\label{43}
\end{equation}
We note that Eqs.\ (\ref{39}) and (\ref{43}) lead to the system of two independent equations: 
\begin{equation}
a^2=\frac{5\gamma}{5\gamma-8\nu},~~~a=\frac{10\gamma-8\nu}{\nu(5\gamma-8\nu)},
\label{44}
\end{equation}
which yield an equation for the parameters $\gamma$ and $\nu$ as
\begin{equation}
(5\gamma-4\nu)^2=\frac{5}{4}\gamma\nu^2(5\gamma-8\nu).
\label{45}
\end{equation}
This equation can also be written in an explicit form:
\begin{equation}
\left(5\gamma-\frac{16\mu}{\sigma}\right)^2=\frac{20\gamma\mu^2}{\sigma^2}\left(5\gamma-\frac{32\mu}{\sigma}\right).
\label{46}
\end{equation}

The variational method developed in this section assumes the minimal mass $m$ or parameter $\mu$ for the trial functions given in Eq.\ (\ref{27}). In this method the mass is a function of the parameter $\sigma$ and hence the minimal mass can be found from
\begin{equation}
\left[\frac{d}{d\sigma}m(\sigma)\right]_{\sigma=\sigma_1}=0.
\label{47}
\end{equation}

We now define $m_1=m(\sigma_1)$ and $\mu_1=m_1c/\hbar$ and also introduce the parameters $\gamma_1$ and $\nu_1$ as
\begin{equation}
\gamma_1=\frac{G_lm_1^n}{\hbar c},~~~\nu_1=\frac{4\mu_1}{\sigma_1}.
\label{48}
\end{equation}
Eq.\ (\ref{46}) and its derivative 
can be written at $\sigma=\sigma_1$ in the form:
\begin{equation}
(5\gamma_1-4\nu_1)^2=\frac{5}{4}\gamma_1\nu_1^2(5\gamma_1-8\nu_1),
\label{49}
\end{equation}
\begin{equation}
5\gamma_1-4\nu_1=-\frac{5}{16}\gamma_1\nu_1(5\gamma_1-8\nu_1)+\frac{5}{4}\gamma_1\nu_1^2,
\label{50}
\end{equation}
where the condition given by Eq.\ (\ref{47}) has been used.
The system of nonlinear equations given by Eq.\ (\ref{49}) and (\ref{50}) has a unique 
solution with $\sigma_1>0$ and $\nu_1>0$:
\begin{equation}
\nu_1=\sqrt{2+2\sqrt{5}},~~~\gamma_1=\frac{4}{5}\nu_1+\frac{1}{5}\nu_1^3.
\label{51}
\end{equation}
Eq.\ (\ref{40}) yields two solutions: $a=\sqrt{2+\sqrt{5}}$ and $a=\sqrt{\sqrt{5}-2}$ for positive and negative sign respectively. We also note that Eq.\ (\ref{43}) requires the inequality $a^2>1$. Thus we have the unique solution $a=\sqrt{2+\sqrt{5}}$ because in the second case the inequality $a^2>1$ is not satisfied.

The criterion for a local minimum of the twice-differentiable function $m(\sigma)$ at the critical point $\sigma_1$ (i.e.\ $m'(\sigma_1)=0$) can be written $m''(\sigma_1)>0$. Eq.\ (\ref{46}) yields the second derivative $\mu''(\sigma_1)=cm''(\sigma_1)/\hbar$ at the critical point $\sigma_1$ as
\begin{equation}
\frac{(15\gamma_1-8\nu_1)\sigma_1^{-1}}{16(n-1)+5n\gamma_1\nu_1-20(n-1)\gamma_1\nu_1^{-1}-4(n+1)\nu_1^2},
\label{52}
\end{equation}
where $n$ is an arbitrary parameter in Eq.\ (\ref{22}). 

One can find that the denominator in Eq.\ (\ref{52}) is zero when $n=0$ (because $5\gamma_1=4\nu_1+\nu_1^3$) and $\mu''(\sigma_1)$ is positive for $n=1,2$. Thus our choice of the exponential parameter as $n=2$ (see Eq.\ (\ref{5}) and (\ref{20})) is consistent with the minimal mass conditions:  $m'(\sigma_1)=0$ and $m''(\sigma_1)>0$. Moreover, the mass defined by minimal mass conditions with $n=2$ has a global minimum at $\sigma=\sigma_1$ because Eqs.\ (\ref{49}) and (\ref{50}) have a unique solution (there is a unique critical point).

We can identify the minimal mass as the mass of the electron (i.e.\ $m_1=m_e$);  
Eqs.\ (\ref{48}) and (\ref{51})  then yield
\begin{equation}
\sigma_1=\frac{4}{\sqrt{2+2\sqrt{5}}}\left(\frac{m_ec}{\hbar}\right).
\label{53}
\end{equation}
The coefficients $A$ and $B$ in the trial functions, namely in Eq.\ (\ref{27}), follow from Eq.\ (\ref{41}) as
\begin{equation}
A=\pm\frac{\sigma_1^{3/2}}{\sqrt{2(a^2+1)}}=\pm\frac{\sqrt{6-2\sqrt{5}}}{4}\sigma_1^{3/2},
\label{54}
\end{equation}

\begin{equation}
B=\mp\frac{a\sigma_1^{3/2}}{\sqrt{2(a^2+1)}}=\mp\frac{\sqrt{2+2\sqrt{5}}}{4}\sigma_1^{3/2}.
\label{55}
\end{equation}
This solution also yields the relation $\tilde{\rm U}(r)=-(5\alpha/8a)\tilde{\Theta}(r)$.
Thus we have the inequality $\tilde{\rm U}(r)\ll |\tilde{\Theta}(r)|$ which demonstrates that in Eqs.\ (\ref{23}) and (\ref{24}) the potential $\tilde{\rm U}(r)$ can be neglected with a high accuracy.

Note that one can choose the coefficients $A$ and $B$ in Eqs.\ (\ref{54}) and (\ref{55}) with upper or lower signs describing the same bound state of the electron. It follows from Eqs.\ (\ref{23})--(\ref{26}) that if $u$ and $v$ are a solution then $-u$ and $-v$ is also a solution of the above equations that differs by a phase transformation: $u\rightarrow e^{i\pi}u$ and $v\rightarrow e^{i\pi}v$.

\section{Bound electron state}

In this section we consider the inner structure of the electron, which is identified in our model with 
the ground state solution. It is given by the trial functions in Eq.\ (\ref{27}) with the parameters defined in Eqs.\ (\ref{53})--(\ref{55}). This solution leads to a unique self-bound electron state
\begin{equation}
u(r)=\pm\frac{2}{\sqrt{\nu_2}}\left(\frac{m_ec}{\hbar}\right)^{3/2}\exp\left[-\left(\frac{2m_ec}{\nu_1\hbar}\right)r\right],
\label{56}
\end{equation}
\begin{equation}
v(r)=\mp\frac{2}{\sqrt{\nu_1}}\left(\frac{m_ec}{\hbar}\right)^{3/2}\exp\left[-\left(\frac{2m_ec}{\nu_1\hbar}\right)r\right],
\label{57}
\end{equation}
where $\nu_1=\sqrt{2+2\sqrt{5}}$ and $\nu_2=\sqrt{58+26\sqrt{5}}$.
We assume that the exponential parameter is $n=2$ and hence the electron mass $m_e=m_1$ given by Eq.\ (\ref{48}) can be written 
\begin{equation}
m_e=q\sqrt{\frac{\hbar c}{G_l}},
\label{58}
\end{equation}
where $q=\sqrt{\gamma_1}= 2.308$. It follows from this relation that the coupling constant is $G_l=2.03\cdot 10^{38}~{\rm cm^3g^{-1}s^{-2}}$. We note that the dimension of the coupling parameter $G_l$  is the same as Newton's gravitation constant $G_N$. 

Furthermore, we introduce the normalized probability density $W(r)$ which is given by 
\begin{equation}
W(r)=\frac{1}{4\pi}[u(r)^2+v(r)^2]=\frac{\sigma_1^3}{8\pi}\exp(-\sigma_1 r).
\label{59}
\end{equation}
The normalization condition has the form $\int WdV=1$ where $dV=4\pi r^2dr$. Thus the $s$-th moment of the probability density is 
\begin{equation}
\langle r^s \rangle=\int r^sW(r)dV=\frac{(s+2)!}{2}\left(\frac{\nu_1}{4}\right)^s\lambda_e^s,
\label{60}
\end{equation}
where $s$ is the integer number and $\lambda_e=\hbar/(m_ec)$ is the Compton wavelength of the electron. This equation yields the electron `radius' as $r_e=\langle r \rangle=1.91\lambda_e$.

We also can write the $\Theta$-field given by Eqs.\ (\ref{29}) and (\ref{30}) in explicit form: 
\begin{equation}
\Theta(r)=k\left[e^{-\sigma_1 r}-\frac{2}{\sigma_1r}(1-e^{-\sigma_1 r}) \right]m_ec^2,
\label{61}
\end{equation}
where $k=4\nu_3/5$ and $\nu_3=\sqrt{6+2\sqrt{5}}$. 
This field has the asymptotic behavior that $\Theta(r)\propto -1/r$ when $\sigma_1 r\gg 1$ and the average $\Theta$-field is 
\begin{equation}
\langle \Theta \rangle=\int \Theta(r) W(r)dV=-\frac{\nu_3}{2}m_ec^2.
\label{62}
\end{equation}
This relation can also be written as
\begin{equation}
\langle \Theta \rangle=-\Gamma\frac{ G_lm_e^2}{\langle r \rangle},
\label{63}
\end{equation}
where $\Gamma=15/(8\nu_3)=0.5794$. This relation has a form similar to the 
gravitational potential of a particle with the mass $m_e$.

The radial distributions for electron charge and mass can be written as $\rho_e(r)=ew(r)$ and $\rho(r)=m_ew(r)$ where the radial distribution of the electron density is 
\begin{equation}
w(r)=4\pi r^2W(r)=\frac{1}{2}\sigma_1^3r^2\exp(-\sigma_1 r).
\label{64}
\end{equation}
This equation yields the conserved charge and mass 
\begin{equation}
\int_0^{+\infty}\rho_e(r)dr=e,~~~\int_0^{+\infty}\rho(r)dr=m_e.
\label{65}
\end{equation}

We plot in Fig. 1 the dimensionless radial distribution of the electron density  $P_e=w/\sigma_1$ as a function of the dimensionless radius $R=\sigma_1 r$. The radial distribution $P_e(R)$ has a maximum at $R=2$ and the average radius is $\langle R \rangle=3$. 

The dimensionless $\Theta$-field can be defined as $\Theta_e=\Theta/(m_ec^2)$. We plot this function  in Fig. 2. The $\Theta_e$ field has the parabolic form $\Theta_e=-k+(k/6)R^2$ for $R\ll 1$ and $\Theta_e=-2k/R$ for $R\gg 1$. 

We emphasize that the scalar $\Theta$-field differs from the potential ${\rm U}$ in sign. It is also shown in Sec. 3 that one can neglect the potential ${\rm U}$ because ${\rm U}\ll |\Theta|$.
The bounded electron state is described by spinors as in Eq.\ (\ref{11}), where the functions $u(r)$ and $v(r)$ in the variational approach are given by Eqs.\ (\ref{56}) and (\ref{57}).

The system of Eqs.\ (\ref{17})--(\ref{21}) or  Eqs.\ (\ref{23})--(\ref{26}) with the normalization condition given by Eq.\ (\ref{21}) can be solved more accurately by numerical methods. In this case one may find the coupling constant $G_l$ assuming that the eigenvalue of the numerical ground state solution coincides with the known mass $m_e$ of electron. Using the value of the constant $G_l$ obtained this way, one can find the eigenvalue $m_2$ of the first excited state, which is the mass of the $\mu$ lepton, and the eigenvalue $m_3$ of the second excited state, which is the mass of the $\tau$ lepton.

\begin{figure}
\centering{}
  \includegraphics[width=9cm]{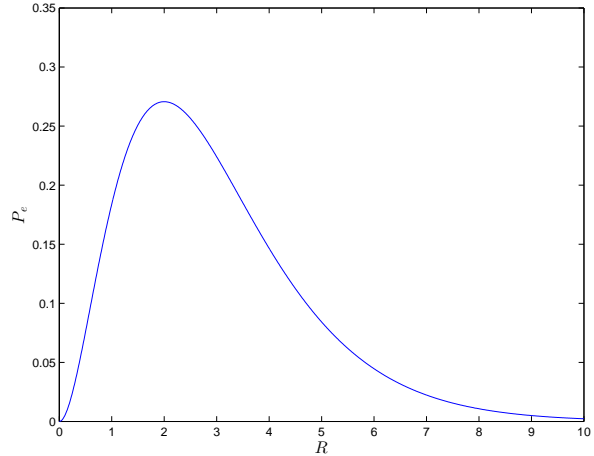}
  \caption{Dimensionless radial distribution of electron density $P_e=(R^2/2)\exp(-R)$ given as function of radius $R=\sigma_1r$.}
\end{figure}

\begin{figure}
\centering{}
  \includegraphics[width=9cm]{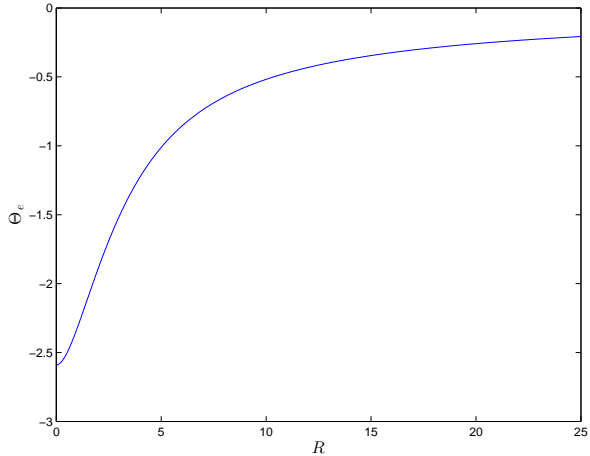}
  \caption{Dimensionless field $\Theta_e=k[e^{-R}-2R^{-1}(1-e^{-R})]$ given as function of radius $R=\sigma_1r$.}
\end{figure}

\section{Conclusions}

We have presented in this Letter a model describing the lepton masses as the eigenvalues of relativistic nonlinear field equations. The eigenfunctions correspond in the model to lepton states with inner structure. In particular, the self-interaction leads to the bound state of electron. Thus in this approach the electron is not a point-like particle as in QED and SM.
The variational solution is found by trial functions given in Eq.\ (\ref{27}) with minimization of the electron mass. The analytical solution yields the coupling constant $G_l$ of the self-interaction in the form $m_e\simeq 2.3/\sqrt{G_l}$. This solution also leads to a finite `radius' of the electron which can be written as $r_e\simeq1.9/m_e$ ($\hbar=c=1$). 

The relation $d\simeq 1/m$ between the size and mass of particles is commonly assumed. We emphasize that the model presented here leads to this relation without any additional assumption. The relation between radius and mass of electron is calculated in Eq.\ (\ref{60}) with $s=1$ using the variational ground state solution of the system of nonlinear integro-differential Eqs.\ (\ref{17}-\ref{21}). Thus this model is able to explain the above phenomenological relation using only these equations. 

There are also the alternative models and direct tests of QED in high-energy electron-positron collisions that confirmed the absence of lepton structures in processes probing distances as small as $R\simeq 10^{-16}~{\rm cm}$, which is much less than its Compton wavelength $\lambda_e\simeq 4\cdot 10^{-11}~{\rm cm}$ (see Ref. \cite{Brod} and numerous references there). 
The simplest alternative may be that the leptons are point-like elementary particles. 
However, this is not necessarily inconsistent with the model presented here,
because in high-energy electron-positron collisions the effective length can be reduced considerably.
 
Now we discuss the model in the context of the SM \cite{SM}, which leads to a physical interpretation of the Higgs field as a `residual' field of the $\Theta$-field after generation the lepton masses. 
 An important property of the Higgs particle \cite{Higgs} is that it interacts with or couples to elementary particles proportionally to their masses. The Higgs boson has self-interactions and the magnitude of quadratic self-interactions is also proportional to the Higgs boson mass.

Note that the vacuum expectation value $v$ is fixed in terms of the Fermi constant determined from muon decay as $v=(\sqrt{2}G_\mu)^{-1/2}$ where $v\simeq 246$ ${\rm GeV}$. The approximate equation $M_H\simeq v/2$ is satisfied between the Higgs boson mass $M_H$ and the vacuum expectation value $v$ because $M_H=v\sqrt{2\lambda}$ and $\lambda\simeq 1/8$ (the Higgs mass is about $M_H\simeq 126$ ${\rm GeV}$ \cite{Higgsm}). Thus we can write the relations: 
\begin{equation}
m_e\simeq \frac{1}{\sqrt{G_l}},~~~M_H\simeq \frac{1}{\sqrt{G_\mu}},
\label{66}
\end{equation}
where the first relation follows from Eq.\ (\ref{58}). 

Hence the Fermi constant $G_\mu$ is connected to the Higgs boson mass $M_H$ by the same relation as the lepton coupling constant $G_l$ is connected to the electron mass $m_e$. 
These relations also suggest that the mass $M_H$  is not a completely free parameter but must be consistent with other known theoretical constrains \cite{Ab,Abd}. 
The first relation in Eq.\ (\ref{66}) follows from the model presented here; the second one has no direct theoretical basis, but is confirmed by recent measurements of the Higgs boson mass \cite{Higgsm}.

The connection between the Higgs scalar field $H$ and the real scalar $\Theta$-field in presented model can be explained by the analogy with the `residual' and inner electrical field in neutral atoms. The $\Theta$-field which yields the bound states of the leptons for different generations is the analog of the inner electrical field forming the bound neutral atoms. The `residual' electrical field of the atoms yields the Lennard-Jones potential describing the interaction between neutral atoms. One can consider the Higgs field $H$ which leads to interaction or coupling of the elementary particles as the analog of the `residual' electric field for interacting neutral atoms. 

To see this, consider the Yukawa fermion Lagrangian which is a part of the full Lagrangian of the SM:
\begin{equation}
{\cal L}_f=\sum_{k=1}^3{\cal L}_f^{k},~~~{\cal L}_f^{k} =-m_k\left(1+\frac{H}{v}\right)\bar{\Psi}_k\Psi_k+...~.
\label{67}
\end{equation}
Here $H$ is the Higgs scalar field and, for example, for the first generation we have $\bar{\Psi}_1\Psi_1=(\bar{e}_Le_R+\bar{e}_Re_L)$.
This Lagrangian also includes the left-handed and right-handed chiral quarks for each generation of leptons. 

We denote the residual scalar field of the $\Theta$-field after the $k$-th fermion generation as $\theta_k$. The fermion Lagrangian describing the residual scalar field after $k$-th fermion generation follows from the replacements $\Theta\rightarrow \theta_k$, $m\rightarrow m_k$ and $\Psi\rightarrow \Psi_k$ in the second term of Eq.\ (\ref{1}). This leads to a fermion Lagrangian given by
\begin{equation}
{\cal L}_f^{'k} =-(m_k+\theta_k)\bar{\Psi}_k\Psi_k,
\label{68}
\end{equation}
where $\bar{\Psi}_k\Psi_k=\bar{\Psi}_{kL}\Psi_{kR}+\bar{\Psi}_{kR}\Psi_{kL}$.

The comparison of Eq.\ (\ref{67}) and (\ref{68}) yields the equation $\theta_k=(m_k/v)H$. 
This is a reasonable relation, because the Higgs particle interacts with or couples to elementary particles proportionally to their masses. 

Thus the Higgs scalar field is equivalent for the $k$-th generation to the residual field $\theta_k$. We note that the description of the residual field of particles by the Higgs scalar field given by $H=v\theta_k/m_k$ is more convenient because it does not depend on the generation number.

\section*{Acknowledgments}

The author is grateful to M. J. Collett for helpful discussion of the results of this work.

\end{document}